\documentclass[10pt,conference]{IEEEtran}
\IEEEoverridecommandlockouts
\usepackage{amsmath,amssymb,amsfonts}
\usepackage{algorithmic}
\usepackage{graphicx}
\usepackage{textcomp}
\usepackage{multirow}
\usepackage{verbatim}
\usepackage{xcolor}
\usepackage{threeparttable}
\usepackage{framed}
\usepackage[numbers]{natbib}
\usepackage{balance}

\def\BibTeX{{\rm B\kern-.05em{\sc i\kern-.025em b}\kern-.08em
    T\kern-.1667em\lower.7ex\hbox{E}\kern-.125emX}}

\colorlet{shadecolor}{gray!10}
\colorlet{framecolor}{black}
\newenvironment{frshaded}{%
	\MakeFramed {\FrameRestore}}%
{\endMakeFramed}
\begin{document}

\title{Towards Understanding the Impacts of Textual Dissimilarity on Duplicate Bug Report Detection}

\author{
\IEEEauthorblockN{Sigma Jahan}
\IEEEauthorblockA{\textit{Department of Computer Science} \\
\textit{Dalhousie University}\\
Nova Scotia, Canada \\
sigma.jahan@dal.ca}
\and
\IEEEauthorblockN{Mohammad Masudur Rahman}
\IEEEauthorblockA{\textit{Department of Computer Science} \\
\textit{Dalhousie University}\\
Nova Scotia, Canada \\
masud.rahman@dal.ca}
}

\maketitle

\begin{abstract} 
About 40\% of software bug reports are duplicates of one another, which pose a major overhead during software maintenance. Traditional techniques often focus on detecting duplicate bug reports that are textually similar. However, in bug tracking systems, many duplicate bug reports might not be textually similar, for which the traditional techniques might fall short. In this paper, we conduct a large-scale empirical study to better understand the impacts of textual dissimilarity on the detection of duplicate bug reports. First, we collect a total of 92,854 bug reports from three open-source systems and construct two datasets containing textually similar and textually dissimilar duplicate bug reports. Then we determine the performance of three existing techniques in detecting duplicate bug reports and show that their performance is significantly poor for textually dissimilar duplicate reports. Second, we analyze the two groups of bug reports using a combination of descriptive analysis, word embedding visualization, and manual analysis. We found that textually dissimilar duplicate bug reports often miss important components (e.g., expected behaviors and steps to reproduce), which could lead to their textual differences and poor performance by the existing techniques. Finally, we apply domain-specific embedding to duplicate bug report detection problems, which shows mixed results. All these findings above warrant further investigation and more effective solutions for detecting textually dissimilar duplicate bug reports.
\end{abstract}


\section{Introduction}

Software bugs are human-made errors in the code that prevent software from working correctly. During software maintenance, software bugs are submitted to a bug-tracking system as \emph{bug reports} \cite{r50}. Hundreds of bugs are reported every day in large software systems (e.g., Mozilla, Eclipse) \cite{r50}. Duplicated bug reports occur when multiple persons submit multiple bug reports for the same bug. Due to the asynchronous nature of bug report submission, traditional bug tracking systems (e.g., Bugzilla) can not prevent duplicate bug reports. Thus, on average, 35.8\%--41.6\% of bug reports remain duplicates of one another in the bug tracking systems \cite{zou2018practitioners}. These duplicate bug reports pose a major overhead during software maintenance since they often cost valuable development time and resources \cite{jalbert2008automated}. 

Manually examining hundreds of bug reports for duplicates is neither feasible nor practical. One of the major challenges in detecting duplicate bug reports is their unstructured and ambiguous nature. Bug reports are written in natural language texts and thus may contain different words describing the same issue. The probability of two persons using the same text to explain the same issue is very low (e.g., 10\%--15\%) \cite{vocaprob}. Given all these inherent challenges, automated detection of duplicate bug reports has become an active research topic since the last decade, which is also known as \emph{bug deduplication} \cite{r25}.

To automate the process of detecting duplicate bug reports, researchers employ various methodologies, including Natural Language Processing (NLP) \cite{r11, r21, sureka2010detecting}, Information Retrieval (IR) \cite{r12, aggarwal2017detecting, alipour2013contextual, r42}, and Machine Learning (ML) \cite{r43, r29, r45, he2020duplicate, budhiraja2018dwen}. However, they are far from perfect due to the complexity and ambiguity of natural language texts. NLP-based techniques might be limited in detecting duplicate reports when there is a textual mismatch between the reports \cite{gupta2021systematic}. IR-based approaches suffer from the \emph{vocabulary mismatch problem} \cite{r51,vocaprob}, a typical phenomenon that stems from two textual documents describing the same concept with different vocabularies. On the other hand, ML-based approaches suffer from data imbalance problems, and a lack of generalizability \cite{gupta2021systematic, r43, ho2020extensions}.

\begin{figure}[htbp]
	\includegraphics[width= 3.5in]{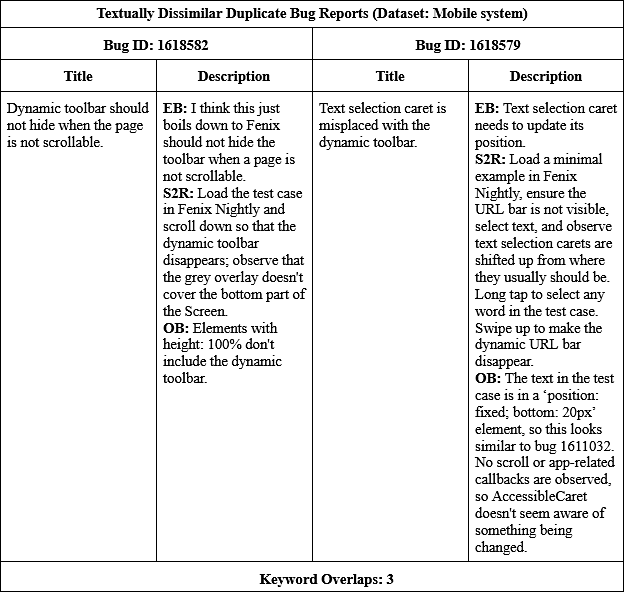}
	\vspace{-.4cm}
	\caption{Textually dissimilar duplicate bug reports from Mobile System}
	\vspace{-.4cm}
	\label{fig:text_dissimilar}
\end{figure}

Duplicate bug reports can be divided into two different categories: those that describe the same issue with similar texts and those that describe two similar issues using different texts (e.g., Fig. \ref{fig:text_dissimilar}) \cite{r11}. The second category refers to duplicate bug reports that have the same underlying root cause but completely different writing styles. There could be instances where two bugs have different observable behaviors (OB) and steps to reproduce (S2R), but they share the same underlying cause. We call these types of duplicate bug reports \textit{textually dissimilar} duplicate bug reports in this paper. According to our investigation, 19\%--23\% of the duplicate bug reports could be textually dissimilar.
 
Most of the existing NLP and IR-based techniques focus on detecting duplicate bug reports that use similar texts. Unlike NLP and IR-based techniques, ML-based techniques can capture the non-linear relationships between two items \cite{obulesu2018machine, almeida2002predictive, bitvai2015non}, and thus have the potential to tackle the challenge of textually dissimilar duplicate bug reports. However, they also suffer from poor outlier handling, class imbalance problem, and a lack of monitoring \cite{deng2018deep}. Thus, automated detection of duplicate bug reports still remains a highly challenging problem that warrants further investigation \cite{r7}.
In this paper, we conduct a large-scale empirical study to better understand the impacts of textual dissimilarity on the detection of duplicate bug reports. First, we collect a total of 92,854 bug reports from three large-scale software systems (Eclipse, Firefox, and Mobile) and \emph{empirically} show how the existing techniques (BM25 \cite{r12}, LDA+GloVe \cite{r13}, and Siamese CNN \cite{r14}) perform poorly in detecting textually dissimilar duplicate bug reports. Second, we compare textually similar and textually dissimilar duplicate bug reports using a \emph{combination} of quantitative and qualitative analyses. We found that the textually dissimilar duplicate bug reports differ from textually similar duplicate bug reports in terms of their underlying semantics and structures. For instance, textually dissimilar duplicate bug reports often have \emph{missing components} or components (e.g., observed behaviors) that are written differently, which could lead to their overall textual differences. Finally, being inspired by the previous successes of domain-specific embedding \cite{chen2021domain, nooralahzadeh2018evaluation}, we apply \emph{domain-specific embedding} to counteract the impact of textual dissimilarity in duplicate bug report detection. We thus answer three important research questions in our study as follows.
\begin{enumerate}
\itemsep3pt
\item[(a)] \textbf{RQ$\mathbf{_1}$: Does the performance of existing techniques differ significantly in duplicate bug report detection between textually similar and textually dissimilar duplicate bug reports?}
\\
We conducted experiments on our dataset using three existing techniques in duplicate bug report detection that employ Information Retrieval (IR), Topic Modeling, and Machine Learning (ML), respectively. We found that the performance of existing techniques is higher (e.g., recall rate@100) is higher (10.02\%--18.45\% for BM25, 2.00\%--6.49\% for LDA+GloVe) in detecting \emph{textually similar} duplicate bug reports than that of \emph{textually dissimilar} duplicate bug reports. Our statistical tests (e.g., Wilcoxon Signed Rank test \cite{woolson2007wilcoxon}, Cliff's delta) also report that their performance is significantly low for textually dissimilar duplicate bug reports. Although our findings reinforce a common belief about existing techniques, they also substantiate it with \emph{solid empirical evidence} of the performance gap between the two categories of duplicate bug reports.

\item[(b)] \textbf{RQ$\mathbf{_2}$: How do textually similar and textually dissimilar duplicate bug reports differ in their semantics and structures?}
\\
To investigate the differences between textually similar and textually dissimilar duplicate bug reports, we use three different analyses: descriptive analysis, embedding analysis, and manual analysis. We found negative skewness in the similarity scores of textually dissimilar duplicate bug reports, which indicates a low textual similarity between each pair. We also visualize their embedding matrices using t-SNE \cite{r32}, a non-linear dimensionality reduction technique. The visualization shows that textually dissimilar duplicate bug reports have a lower dimensional space than that of textually similar duplicate bug reports, which indicates a lower pairwise distance within the embedding space (Fig. \ref{fig:t-sne}). Finally, our manual analysis suggests that textually dissimilar duplicate bug reports often \emph{miss important components} (e.g., steps to reproduce) or they have components (e.g., observed behaviors) that are written differently, which could lead to their overall textual differences. 

\item[(c)] \textbf{RQ$\mathbf{_3}$: Does domain-specific embedding help improve the detection of textually dissimilar duplicate bug reports?
}
\\
Our experiments in RQ$_1$ use a pre-trained, generic embedding model, GloVE \cite{r54}, which might not have effectively overcome the challenges of textual dissimilarity in duplicate bug report detection \cite{efstathiou2018word, li2020advantages, rezaeinia2019sentiment}. We thus retrain our selected DL-based technique (e.g., Siamese CNN \cite{r14}) with a domain-specific embedding model \cite{chen2021domain, roy2017learning, nooralahzadeh2018evaluation} and repeat our experiments. In particular, we analyze 92,854 bug reports from Eclipse, Firefox, and Mobile systems to capture domain-specific embedding, which is then used to retrain the DL-based technique. We have also used oversampling to deal with imbalanced data problems during model training \cite{yap2014application}. We found that domain-specific embedding shows \emph{mixed results} by improving the detection of textually dissimilar duplicate bug reports but worsening the detection of textually similar duplicate bug reports.
\end{enumerate}

\section{Study Methodology}\label{sec:methodology}

\subsection{Construction of dataset}\label{sec:dataset-construction}
\textbf{Dataset collection.} 
We collect duplicate bug reports from three large-scale, open-source software systems -- Eclipse, Firefox, and Mobile -- using a popular bug tracking system, namely \emph{Bugzilla}. Existing studies \cite{gupta2021systematic,alipour2013contextual,r43,r45} have frequently used these systems, which makes them suitable for our research. Besides, these systems have stemmed from diverse application domains. Eclipse is a popular open-source Integrated Development Environment (IDE) written in Java. Firefox is a popular open-source web browser. While the above two systems are desktop-based applications, the remaining systems are mobile-based (e.g., Firefox for iOS, Focus for iOS, and GeckoView for Android). For the sake of brevity, we combine these three small systems and call them \emph{Mobile} in the rest of the paper. Almost all the existing studies on duplicate bug report detection used the dataset dated before 2017 \cite{gupta2021systematic}, which might not be an ideal representation of recent software bugs and issues \cite{r28}. In order to avoid the issue of concept drift \cite{lu2018learning}, we choose the bug reports from the last five years. We thus collected a total of 92,854 bug reports from three systems that were submitted within the last five years (01-01-2017 to 01-01-2022).

For many duplicate bug reports, the master reports were created before the last five years (Eclipse: 766, Firefox: 3514, Mobile: 214). In order to make a complete dataset with all the duplicate bug reports along with their master bug reports, we also retrieved them separately. Table \ref{tab:dataset} summarizes our study dataset. We see that about 6.60\%, 20.53\%, and 10.56\% of the submitted bug reports were duplicates in Eclipse, Firefox, and Mobile systems, respectively. 

\setlength{\arrayrulewidth}{0.1mm}
\setlength{\tabcolsep}{11pt}
\renewcommand{\arraystretch}{1.4}
\begin{table}[!t]
\centering
\caption{Study Dataset}
\label{tab:dataset}
\resizebox{3.5in}{!}{
			\begin{threeparttable}
\begin{tabular}{|l|c|c|c|}
\hline
\textbf{Dataset (2017 -- 2022)} & \textbf{Eclipse} & \textbf{Firefox} & \textbf{Mobile} \\ \hline
\hline
\textbf{Whole Dataset} & 49,244 & 38,290 & 5,320 \\ \hline
\textbf{Total Duplicate} & 3,248 & 7,859 & 562 \\ \hline
\textbf{Duplicate Ratio} & 6.60 \% & 20.53\% & 10.56\%  \\ \hline
\multicolumn{4}{c}{\textbf{Experimental Dataset (BM25, LDA + GloVe)}} \\
\hline
\textbf{Textually Similar Duplicate} & 679 & 1,414 & 122 \\ \hline
\textbf{Textually Dissimilar Duplicate} & 662 & 1,455 & 131 \\ \hline
\multicolumn{4}{c}{\textbf{Experimental Dataset (Siamese CNN)}} \\
\hline
\textbf{Training Set} & 39,395 & 30,632 & 4,256 \\ \hline
\textbf{Testing Set} & 9,848 & 7,658 & 1,064 \\ \hline
\textbf{Textually Similar Duplicate} & 504 & 610 & 117 \\ \hline
\textbf{Textually Dissimilar Duplicate} & 497 & 734 & 89 \\ \hline
\end{tabular}
\end{threeparttable}}
\vspace{-.5cm}
\end{table}

\textbf{Data cleaning and preprocessing.} We capture four key fields from each bug report for our study: (a) bug id, which is unique for each bug report, (b) duplicate bug id, which points to the duplicate bug report, (c) title and description of the bug report, and (d) resolution, which indicates the duplicate status of a bug report.

We collect \emph{title} and \emph{description} from each bug report since they capture pertinent information for detecting duplicate bug reports. We clean and preprocess the \emph{title} and \emph{description} 
from each bug report using several steps as follows. 

We apply standard natural language preprocessing to the title and description texts. First, we remove stopwords using a standard set of stopwords, which have little to no significance in capturing semantics. Then we perform token splitting along with the removal of punctuation marks,  non-alphanumeric characters, numbers, HTML meta tags, and URLs. We also replace any non-alphanumeric characters with spaces and transform the text into lowercase \cite{r25}. Lastly, to transform each term into its base form, we use lemmatization using the NLTK library in Python. As performed by an earlier work \cite{r25}, we discard any description with fewer than 50 characters since they do not contain enough information to be meaningful. On the other hand, bug reports might have long description text containing source code, lengthy stack traces, and log files, which could be noisy \cite{r25}. Hence, several previous studies \cite{r14, r25} selected bug reports containing at most 350 to 500 tokens. We experimented with 350, 500, and 1000 tokens. Using 500 tokens, the model delivers the best performance. Thus, we chose 500 as our token limit for the bug reports.

\textbf{Construction of triplets.} After we clean and preprocess the dataset, we construct triplets (b, b+, b-) where the existing techniques are supposed to detect b+. Here, b means the query bug report, b+ means duplicate bug report, and b- means non-duplicate bug reports. We created these triplets inspired by an existing work \cite{r14}. In the existing work, the duplicates (b,b+) were determined based on bug-tracking systems, whereas non-duplicates (b,b-) were randomly selected from the dataset. We also follow the same process in our ground truth construction. Then (b, b+) pairs were used as the ground truth for evaluating the existing techniques in duplicate bug report detection.\\
\indent
\textbf{Dataset preprocessing for ML-based approach.}
To design an ML-based model (e.g., Siamese CNN) for duplicate bug report detection, pairwise bug reports containing texts and ground truth are required. Hence, we extract (b, b+) and (b, b-) pairs from the triplets above to generate our positive and negative samples, respectively. A similar process was adopted by the existing literature \cite{r14}. For training and testing, we split the whole dataset into an 80:20 ratio with random shuffling. Table \ref{tab:dataset} (bottom section) shows our Machine Learning models' training and testing datasets from all three systems.\\
\setlength{\arrayrulewidth}{0.1mm}
\setlength{\tabcolsep}{3pt}
\renewcommand{\arraystretch}{1.9}
\begin{table*}
\centering
\caption{Construction of textually similar and dissimilar duplicate pairs using n-gram based similarity scores}
\vspace{-.2cm}
\label{tab:similarity_score}
\resizebox{5in}{!}{
			\begin{threeparttable}
\begin{tabular}{|cccc|ccc|ccc|cc|}
\hline
\multicolumn{4}{|c|}{\textbf{Unigram}}                                                                                                                                                                      & \multicolumn{3}{c|}{\textbf{Bigram}}                                                                                                                                       & \multicolumn{3}{c|}{\textbf{Trigram}}                                                                                                                                     & \multicolumn{2}{c|}{\textbf{Duplicate Bug Reports}}                                                                                                 \\ \hline
\multicolumn{1}{|c|}{Dataset} & \multicolumn{1}{c|}{Median} & \multicolumn{1}{c|}{\begin{tabular}[c]{@{}c@{}}Lower  \\ Quartile\end{tabular}} & \begin{tabular}[c]{@{}c@{}}Upper   \\ Quartile\end{tabular} & \multicolumn{1}{c|}{Median} & \multicolumn{1}{c|}{\begin{tabular}[c]{@{}c@{}}Lower  \\ Quartile\end{tabular}} & \begin{tabular}[c]{@{}c@{}}Upper  \\ Quartile\end{tabular} & \multicolumn{1}{c|}{Median} & \multicolumn{1}{c|}{\begin{tabular}[c]{@{}c@{}}Lower  \\ Quartile\end{tabular}} & \begin{tabular}[c]{@{}c@{}}Upper \\ Quartile\end{tabular} & \multicolumn{1}{c|}{\begin{tabular}[c]{@{}c@{}}Textually \\ Similar\end{tabular}} & \begin{tabular}[c]{@{}c@{}}Textually \\ Dissimilar\end{tabular} \\ \hline
\hline
\multicolumn{1}{|c|}{Eclipse} & \multicolumn{1}{c|}{0.0502} & \multicolumn{1}{c|}{0.0361}                                                     & 0.0675                                                      & \multicolumn{1}{c|}{0.0261} & \multicolumn{1}{c|}{0.0189}                                                     & 0.0368                                                     & \multicolumn{1}{c|}{0.0212} & \multicolumn{1}{c|}{0.0156*}                                                    & 0.0297*                                                   & \multicolumn{1}{c|}{679}                                                          & 662                                                             \\ \hline
\multicolumn{1}{|c|}{Firefox} & \multicolumn{1}{c|}{0.0633} & \multicolumn{1}{c|}{0.0483}                                                     & 0.0776                                                      & \multicolumn{1}{c|}{0.0298} & \multicolumn{1}{c|}{0.0233}                                                     & 0.0355                                                     & \multicolumn{1}{c|}{0.0232} & \multicolumn{1}{c|}{0.0179*}                                                    & 0.0274*                                                   & \multicolumn{1}{c|}{1414}                                                         & 1455                                                            \\ \hline
\multicolumn{1}{|c|}{Mobile}  & \multicolumn{1}{c|}{0.0679} & \multicolumn{1}{c|}{0.0488}                                                     & 0.0944                                                      & \multicolumn{1}{c|}{0.0451} & \multicolumn{1}{c|}{0.0316}                                                     & 0.0639                                                     & \multicolumn{1}{c|}{0.0403} & \multicolumn{1}{c|}{0.0283*}                                                    & 0.0565*                                                   & \multicolumn{1}{c|}{122}                                                          & 131                                                             \\ \hline
\end{tabular}
\end{threeparttable}}
\vspace{-.3cm}
\end{table*}
\indent
\textbf{Constructing subsets of study datasets based on textual similarity.}
We divide our dataset into textually similar and textually dissimilar duplicate bug reports, which are essential for answering our research questions. 

First, we store all duplicate bug reports as pairs in a separate dataset. Then, we collect N-grams (n = 1, 2, and 3) to compute the textual similarity \cite{buscaldi2012irit} of each duplicate bug report pair.  We use the cosine similarity metric \cite{r11} to calculate the textual similarity between two bug reports. After getting the similarity score between the duplicate pairs, we analyze their descriptive statistics to determine their median similarity score, lower quartile (25th percentile) value, and upper quartile (75th percentile) value. For duplicate pairs that have a similarity score less than the lower quartile value, we denote them as \textit{textually dissimilar} duplicate bug reports.\\
\indent
On the other hand, for duplicate pairs that have a similarity score more than the upper quartile value, we denote them as \textit{textually similar} duplicate bug reports. We repeat all these steps using Unigram, Bigram, and Trigram to collect the common set of textually similar and textually dissimilar duplicate bug reports across three trials for our experiment. Table \ref{tab:similarity_score} shows similarity scores used for various N-grams to construct our textually similar and dissimilar duplicate bug reports. Finally, we got two subsets for textually similar and textually dissimilar duplicate bug reports, respectively (Eclipse: 679 \& 662, Firefox: 1414 \& 1455, Mobile: 122 \& 131). 

\subsection{Replication of existing techniques for experiments}
To answer our first research question, we needed to replicate existing techniques on duplicate bug report detection. We thus select suitable representatives from the frequently used methodologies in duplicate bug report detection. In particular, we choose baseline methods from three frequently used methodologies -- Information Retrieval, Topic Modeling, and Machine Learning. We select BM25 \cite{r12} from Information Retrieval, LDA+GloVe \cite{r13} from Topic-Modeling, and Siamese CNN \cite{r14} from Deep Learning for our experiment. Most of the recent models are based on these three primary approaches with incremental improvements  \cite{gupta2021systematic, bansal2021literature, neysiani2019duplicate}. We chose these baseline methods to determine the impact of textual dissimilarity on duplicate bug report detection without the effect of compounding factors (e.g., severity, priority, components, products, multimedia attachments).

We used the authors' replication package for the LDA+GloVe model \cite{r13}. On the other hand, the replication packages of BM25 \cite{r12}, and Siamese CNN \cite{r14} were not publicly available, and thus those techniques were carefully re-implemented by us based on the corresponding papers.

Information Retrieval (IR) relies on keyword overlaps between any two documents. We select a representative of IR, namely BM25, with default parameters (\emph{k1=1.5}, \emph{b=0.75}) for retrieving the duplicate bug reports. \citet{r12} first use BM25 for duplicate bug report detection. While BM25 is an established technique, it suffers from Vocabulary Mismatch Problem (VMP) \cite{r51, vocaprob}. Several studies \cite{r13, nguyen2012duplicate, r61} adopt Topic Modeling in duplicate bug report detection to overcome this challenge. Latent Dirichlet Allocation (LDA) is a Topic Modeling approach that has the potential to overcome the vocabulary mismatch problem \cite{r58}. We call this approach LDA+GloVe in our research. As done by the original work \cite{r13}, we use LDA for topic-based clustering (topic number=20), GloVe for the pre-trained word embedding (embedding dim = 100), and a unified text similarity measure (cosine similarity and euclidean metrics) for ranking the topmost, similar bug reports against a given bug report.

Unlike the above two techniques, Machine Learning based approaches might be able to find non-linear relationships between dependent and independent variables \cite{obulesu2018machine, almeida2002predictive, bitvai2015non}. We thus implement the Siamese Convolutional Neural Network (CNN) for duplicate bug report detection by adapting an earlier research \cite{r14}. We train our Siamese CNN model using K-fold cross-validation (e.g., \emph{K=10}) and batch gradient descent, where 
original authors' batch sizes (e.g., 512 for Eclipse and Firefox, 256 for Mobile), learning rate (e.g., 0.001) and epoch number (e.g., 12) were chosen during the training phase to avoid model overfitting. Since the Mobile system has a small number of bug reports, a smaller batch size was chosen.

\subsection{Performance evaluation}
As we detected duplicate bug reports using Information Retrieval, Topic Modeling, and Machine Learning techniques in our experiments, they were evaluated using appropriate performance metrics from these domains. In the case of BM25 and LDA+GloVe, we find all the duplicate bug reports for a given query bug report. Then we used Recall-rate@K \cite{gupta2021systematic, r13, r15, r21}, one of the most popular performance metrics, to evaluate the IR and Topic Modeling approaches. On the other hand, for the deep learning-based model (Siamese CNN), we used traditional metrics such as F1 score, AUC, recall, and precision \cite{gupta2021systematic}. We used different evaluation metrics based on the original works \cite{r12, r13, r14}. It should be noted that our main goal was to contrast the performance of existing techniques between two sets of duplicate bug reports rather than to compare the techniques.\\
\indent
\textbf{Recall-rate@K:} Recall-rate@K determines the percentage of bug reports for each of which the duplicate bug report is found within the top K positions \cite{r29}.
\begin{equation}
Recall Rate@K = {N_{detected}}/{N_{total}}
\end{equation}

N$_{detected}$ is the number of bug reports for which the duplicate reports have been correctly detected and N$_{total}$ is the number of total bug reports. We used nine different values of K (K = 1, 5, 10, 20, 25, 30, 50, 75, 100) to calculate the results of our IR-based and Topic Modeling techniques -- BM25 and LDA+GloVe models.
~\\
\indent
\textbf{Precision:} Precision determines the percentage of bug reports for which duplicate bug reports are correctly detected. We calculate the precision of a technique as follows:
\begin{equation}
Precision=\frac{TP}{TP + FP}
\end{equation}
Here TP = true positive and FP = false positive.
\\
\indent
\textbf{Recall:} Recall determines the percentage of all duplicate bug reports that are correctly detected by a technique. We calculate the metrics as follows:
\begin{equation}
Recall=\frac{TP}{TP + FN}
\end{equation}
Here, TP = true positive and FN = false negative.\\
\indent
\textbf{F1-measure:} While both precision and recall focus on a specific aspect of a technique's effectiveness, F1-measure is a more comprehensive and effective method for evaluation. We take the harmonic mean of precision and recall to compute the F1-measure as follows:
\begin{equation}
F_{1}Measure=2\times\frac{Precision * Recall}{Precision + Recall}
\end{equation}
\\
\indent
\textbf{AUC:} The Receiver Operating Characteristics (ROC) curve is a probability curve that separates between true-positive and false-positive rates \cite{r57}. Area Under the Curve (AUC) calculates the fraction of the area that falls under the ROC \cite{r57}. The AUC score ranges between 0 and 1, with 1 indicating that the model can perfectly classify observations into classes. The positive and negative samples are frequently imbalanced in actual data, such as ours. This imbalance significantly impacts precision and recall, whereas Area Under Curve (AUC) is robust against the data imbalance.

After conducting the experiments, we evaluated our BM25 and LDA+GloVe models using Recall-rate@K, as used by existing work \cite{r12, r13}. On the other hand, we have evaluated the Siamese CNN model, as was done by the original work \cite{r14}, using the remaining performance metrics.

\setlength{\arrayrulewidth}{0.1mm}
\setlength{\tabcolsep}{11pt}
\renewcommand{\arraystretch}{1.2}
\begin{table}[!t]
\centering
\caption{Experimental results of IR \& LDA-Based Techniques}
\vspace{-.2cm}
\label{tab:ir-whole-dataset}
\resizebox{3.5in}{!}{
			\begin{threeparttable}
\begin{tabular}{|c|l|c|c|c|c|}
\hline
\textbf{Dataset} & \textbf{Method} & \textbf{k = 1} & \textbf{k = 5} & \textbf{k = 10} &\textbf{k =100} \\ \hline
\hline
\multirow{2}{*} {Eclipse} & {BM25} & 22.64 & 36.60 & 42.03 & 57.31
 \\ \cline{2-6} 
 & {LDA+GloVe} & 0 & 5.5 & 10.5 & 16.0
 \\ \hline
\multirow{2}{*}{{Firefox}} & {BM25} & 16.11 & 27.98 & 33.64 & 52.68 \\ \cline{2-6} 
 & {LDA+GloVe} & 0 & 8.5 & 10.5 & 16.5 \\ \hline
\multirow{2}{*}{{Mobile}} & {BM25} & 17.25 & 28.95 & 35.38 & 57.60 \\ \cline{2-6} 
 & {LDA+GloVe} & 0 & 2.5 & 7.0 & 20.0 \\ \hline
\end{tabular}
\end{threeparttable}
}
\vspace{-.4cm}
\end{table}

\setlength{\arrayrulewidth}{0.1mm}
\setlength{\tabcolsep}{3pt}
\renewcommand{\arraystretch}{1.5}
\begin{table*}
\centering
\caption{Performance of IR \& LDA-based Techniques  with Textually Similar and Dissimilar Duplicate Bug Reports}
\label{tab:ir-similar-dissimilar}
\begin{tabular}{|c|l|cccc|cccc|}
\hline
 &                                   & \multicolumn{4}{c|}{\textbf{\begin{tabular}[c]{@{}c@{}}Textually Similar Duplicates\\ (Recall-rate@k)\%\end{tabular}}}                  & \multicolumn{4}{c|}{\textbf{\begin{tabular}[c]{@{}c@{}}Textually Dissimilar Duplicates\\ (Recall-rate@k)\%\end{tabular}}}                \\ \cline{3-10} 
\multirow{-2}{*}{\textbf{Dataset}} & \multirow{-2}{*}{\textbf{Method}} & \multicolumn{1}{c|}{\textbf{k=1}} & \multicolumn{1}{c|}{\textbf{k=5}} & \multicolumn{1}{c|}{\textbf{k=10}}                 & \textbf{k=100} & \multicolumn{1}{c|}{\textbf{k=1}} & \multicolumn{1}{c|}{\textbf{k=5}} & \multicolumn{1}{c|}{\textbf{k=10}}                 & \textbf{k=100} \\ \hline
\hline
 & {BM25}                     & \multicolumn{1}{c|}{24.15}        & \multicolumn{1}{c|}{37.63}        & \multicolumn{1}{c|}{43.26} & 62.78          & \multicolumn{1}{c|}{21.83}         & \multicolumn{1}{c|}{37.50}        & \multicolumn{1}{c|}{41.27} & 52.58          \\ \cline{2-10} 
\multirow{-2}{*}{{Eclipse}} & {LDA +  GloVe}              & \multicolumn{1}{c|}{0.00}            & \multicolumn{1}{c|}{10.0}         & \multicolumn{1}{c|}{13.5}                          & 20.50           & \multicolumn{1}{c|}{0.00}            & \multicolumn{1}{c|}{6.50}          & \multicolumn{1}{c|}{11.50}                          & 18.50           \\ \hline
& {BM25}                     & \multicolumn{1}{c|}{20.72}        & \multicolumn{1}{c|}{34.49}        & \multicolumn{1}{c|}{39.92}                         & 57.58          & \multicolumn{1}{c|}{11.64}        & \multicolumn{1}{c|}{20.82}        & \multicolumn{1}{c|}{26.56} & 46.72          \\ \cline{2-10} 
\multirow{-2}{*}{{Firefox}} & {LDA +  GloVe}              & \multicolumn{1}{c|}{0.00}            & \multicolumn{1}{c|}{6.00}          & \multicolumn{1}{c|}{8.50}                           & 14.49          & \multicolumn{1}{c|}{0.00}            & \multicolumn{1}{c|}{2.00}          & \multicolumn{1}{c|}{5.00}                           & 8.00            \\ \hline
& {BM25}                     & \multicolumn{1}{c|}{22.00}        & \multicolumn{1}{c|}{44.00}        & \multicolumn{1}{c|}{48.00}                         & 78.00          & \multicolumn{1}{c|}{15.73}        & \multicolumn{1}{c|}{28.09}        & \multicolumn{1}{c|}{35.96}                         & 59.55          \\ \cline{2-10} 
\multirow{-2}{*}{{Mobile}}  & {LDA +  GloVe}              & \multicolumn{1}{c|}{0.00}            & \multicolumn{1}{c|}{4.00}            & \multicolumn{1}{c|}{7.50}                           & 20.50           & \multicolumn{1}{c|}{0.00}            & \multicolumn{1}{c|}{2.50}          & \multicolumn{1}{c|}{4.50}                           & 15.00           \\ \hline
\end{tabular}
\vspace{-.5cm}
\end{table*}

\section{Study Findings} \label{sec:findings}

\subsection{RQ$\mathbf{_1}$: Does the performance of existing techniques differ significantly in duplicate bug report detection between textually similar and textually dissimilar duplicate bug reports?}

We first evaluate the existing techniques against our whole dataset. Tables \ref{tab:ir-whole-dataset} and \ref{tab: domain-specific embedding} summarize their performances. 

From Table \ref{tab:ir-whole-dataset}, we see that the BM25 approach, on average, performs higher with the Eclipse system than with Firefox and Mobile systems. On the other hand, the performance of the LDA+GloVe  model is approximately 38.36\% lower than that of BM25 for Recall-rate@100. LDA is limited in modeling topic correlations \cite{blei2006correlated}, which could be crucial to duplicate bug report detection. Table \ref{tab: domain-specific embedding} (top section) shows that the performance of our ML-based technique in detecting duplicate bug reports ranges from 61.41\% -- 84.80\% in terms of AUC. Precision, Recall, and F1-measure are slightly higher for Eclipse than for Firefox. Firefox has a better AUC score than Eclipse, as the AUC metric is robust to imbalanced data \cite{yap2014application}. Precision, Recall, F1-measure, and AUC scores are slightly higher for the Mobile dataset than for the other two systems. Overall, the ML-based approach delivers the highest performance in duplicate bug report detection.

\begin{figure*}[htbp]
\centering
	\includegraphics[width = 5in]{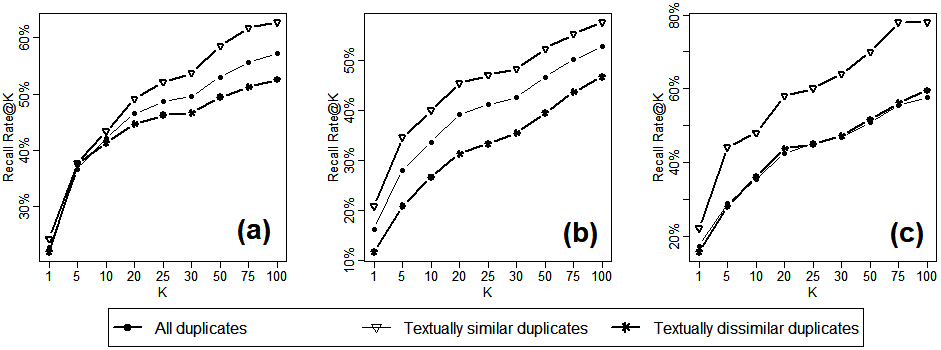}
	\vspace{-.2cm}
	\caption{Performance of BM25 with all textually similar and dissimilar duplicate bug reports from (a) Eclipse, (b) Firefox, and (c) Mobile systems}
	\vspace{-.5cm}
	\label{fig:recall-rate}
\end{figure*}

While the above analysis focuses on the whole dataset, we also determine the performance gap of existing techniques between \emph{textually similar} and \emph{textually dissimilar} duplicate bug reports. Table \ref{tab:ir-similar-dissimilar} shows that BM25 delivers a higher Recall-rate@K for textually similar duplicate bug reports than for dissimilar ones across all three systems - Eclipse, Firefox, and Mobile. For instance, with K=100, the difference between these two sets ranges from 10.20\% to 18.45\%. Fig. \ref{fig:recall-rate} further shows the performance difference between these two sets of bug reports for various K values. We see that the difference is noticeably higher for Firefox and Mobile systems. 

On the other hand, for the LDA+GloVe model, the performance differences between textually similar and textually dissimilar duplicate bug reports are 2.00\% for Eclipse, 6.49\% for Firefox, and 5.5\% for the Mobile system with K=100 (Table \ref{tab:ir-similar-dissimilar}). The performance gap is higher for Firefox and Mobile systems in the same manner as BM25. One possible reason behind this could be the higher duplicate ratios in Firefox and Mobile systems (see Table \ref{tab:dataset}).

Table \ref{tab: domain-specific embedding} shows the performance of our Machine Learning model - Siamese CNN - for both sets of duplicate bug reports across three systems. Here, the performance difference between textually similar and textually dissimilar duplicates is less apparent than that of the traditional methods above. Machine Learning models, especially deep learning models, can capture more contextual information beyond the syntax \cite{lai2015recurrent}, which might explain the phenomenon. From Table \ref{tab: domain-specific embedding}, we see that the performance difference is smaller for Eclipse than for the other two datasets. For instance, the AUC differences between textually similar and textually dissimilar duplicate bug reports are 9.57\% for Eclipse, 9.38\% for Firefox, and 2.97\% for the Mobile dataset. In the case of the F1-measure, the performance difference for the Eclipse dataset is 10.92\%, whereas, for the Firefox dataset, it is 8.48\%. For Mobile data, the performance difference is 10.00\%. We also note that the performance of our ML-based technique is lower for the two subsets of bug reports than for the whole dataset. As shown in Table \ref{tab:dataset}, both subsets contain a smaller number of duplicate bugs than the whole dataset. That is, the two experiments use the same trained model but are tested with different numbers of test samples, which might explain the finding.\\
\indent
We also perform statistical tests to determine the significance of the performance gap between textually similar and dissimilar duplicate bug reports (Table \ref{tab:significance test}). For each of the three systems, we first evaluate BM25 and LDA+GloVe using Recall-rate@K measures against textually similar and dissimilar duplicate bug reports where we consider various K values (K = 1, 5, 10, 20, 25, 30, 50, 75, 100). Then, we performed \emph{Shapiro-Wilk normality test} \cite{royston1992approximating} to determine the distribution of each set. We got two non-normal distribution pairs (LDA+GloVe for Eclipse and Firefox) out of six sets of pairs. Then we used appropriate parametric, non-parametric tests and effect size tests to compare the two sets of Recall-rate@k values from textually similar and dissimilar duplicate bug reports. For the normal distribution, we used \emph{paired t-test} as the parametric test \cite{r31}, and for the non-parametric test, we used the \emph{Wilcoxon Signed‐Rank test} \cite{woolson2007wilcoxon}. In both types of significance tests, the p-values were less than the threshold (0.05) except for two sets (BM25 in the Eclipse system \& LDA+GloVe in the Mobile system). Thus, the null hypothesis can be rejected for all comparisons except for the case of BM25 in Eclipse and LDA+GloVE in the Mobile system. In other words, the performances of BM25 and LDA+GloVE techniques are significantly different between textually similar and dissimilar duplicate bug reports. \\
\indent
While the significance of a result indicates how probable it is that it is due to chance, the effect size indicates the extent of the difference \cite{r37}. Our experiments found different effect sizes ranging from medium to large (Table \ref{tab:significance test}). We see that the effect size of BM25 is large for Firefox and Mobile systems and medium for Eclipse. On the other hand, the LDA+GloVe model has a medium to large effect size across the three systems. Thus, our results from effect size tests reinforce the above finding from significance tests. In other words, the existing techniques perform significantly poorly detecting textually dissimilar duplicate bug reports. Even though our findings above mostly match natural intuition, we performed extensive experiments on three different systems using three different methodologies, which resulted in strong empirical evidence. Thus, we not only reinforce the existing belief about the existing techniques on duplicate bug detection but also substantiate it with solid empirical evidence.

\FrameSep.3em
\begin{frshaded}
	\noindent
	\textbf{Summary of RQ$\mathbf{_1}$:} The performances of existing techniques (e.g., BM25, LDA+GloVe) are \emph{significantly} lower in detecting textually dissimilar duplicate bug reports 
 than that of textually similar duplicate bug reports. Our finding also substantiates a common belief about existing techniques on duplicate bug report detection with \emph{solid empirical evidence}. 
\end{frshaded} 
\setlength{\arrayrulewidth}{0.1mm}
\setlength{\tabcolsep}{11pt}
\renewcommand{\arraystretch}{1.2}
\begin{table}
\centering
\caption{Statistical Tests for the Performance Gap between Textually Similar and Dissimilar Duplicates}
\label{tab:significance test}
\resizebox{3.5in}{!}{
 \begin{threeparttable}
\begin{tabular}{|c|l|c|c|c|}
\hline
\textbf{Dataset} & \textbf{Method} & \textbf{PD} & \textbf{\begin{tabular}[c]{@{}c@{}}Significance\\ (p-value)\end{tabular}} & \textbf{Effect Size} \\ \hline
\hline
\multirow{2}{*}{Eclipse} & BM25 & N & 0.2865   & Medium (0.52) \\ \cline{2-5} 
 & LDA+GloVe & NN & 0.0113*  & Large (0.80) \\ \hline
\multirow{2}{*}{Firefox} & BM25 & N & 0.0329** & Large (1.10) \\ \cline{2-5} 
 & LDA+GloVe & NN & 0.0103*  & Medium (0.31) \\ \hline
\multirow{2}{*}{Mobile} & BM25 & N & 0.0586** & Large (0.96) \\ \cline{2-5} 
 & LDA+GloVe & N & 0.1471   & Medium (0.72) \\ \hline
\end{tabular}
\centering
\textbf{PD}=Probability Distribution, \textbf{NN}=Non-normal, \textbf{N}=Normal, \textbf{*}=Significant, \textbf{**}=Strongly Significant
\end{threeparttable}
\vspace{-.5cm}
}
\end{table}

\setlength{\arrayrulewidth}{0.1mm}
\setlength{\tabcolsep}{11pt}
\renewcommand{\arraystretch}{1.2}
\begin{table}[!t]
\centering
\caption{Descriptive analysis of Similarity Scores between Bug Reports}
\label{tab:descriptive-analysis}
\resizebox{3.5in}{!}{
			\begin{threeparttable}
\begin{tabular}{|l|c|c|c|c|c|}
\hline
\textbf{Dataset} & \textbf{Skew} & \textbf{Kurt} & \textbf{Mean} & \textbf{Median} & \textbf{Std} \\ \hline
\multicolumn{6}{c} {\textbf{Eclipse}} \\
\hline
{\begin{tabular}[l]{@{}c@{}}Textually Similar\end{tabular}} & 0.64 & -0.83 & 0.09 & 0.08 & 0.02 \\ \hline
{\begin{tabular}[c]{@{}c@{}}Textually Dissimilar\end{tabular}} & -0.70 & 0.03 & 0.02 & 0.02 & 0.01 \\ \hline
\multicolumn{6}{c} {\textbf{Firefox}} \\
\hline
{\begin{tabular}[c]{@{}c@{}}Textually Similar\end{tabular}} & 0.89 & 0.37 & 0.09 & 0.09 & 0.01 \\ \hline
{\begin{tabular}[c]{@{}c@{}}Textually Dissimilar\end{tabular}} & -0.50 & -0.67 & 0.03 & 0.03 & 0.01 \\ \hline
\multicolumn{6}{c} {\textbf{Mobile}} \\
\hline
{\begin{tabular}[c]{@{}c@{}}Textually Similar\end{tabular}} & -0.88 & -0.53 & 0.16 & 0.17 & 0.03 \\ \hline
{\begin{tabular}[c]{@{}c@{}}Textually Dissimilar\end{tabular}} & -0.38 & -0.58 & 0.03 & 0.03 & 0.01 \\ \hline
\end{tabular}
\end{threeparttable}}
\vspace{-.4cm}
\end{table}

\subsection{RQ$_2$: How do textually similar and textually dissimilar duplicate bug reports differ in their semantics and structures?}
In this research question, we investigate how textually dissimilar duplicate bug reports might be different from textually similar duplicate bug reports. We answer this question using three different analyses -- descriptive analysis, embedding analysis, and manual analysis -- as follows:

\begin{figure}[htbp]
	\includegraphics[width= 3.6in]{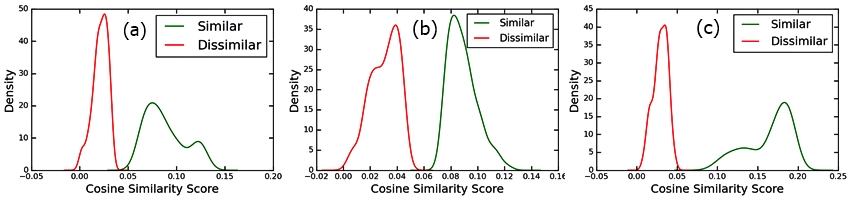}
		\vspace{-.4cm}
	\caption{Distribution of similarity measures for textually similar and dissimilar duplicate bug reports from (a) Eclipse, (b) Firefox, and (c) Mobile system}
		\vspace{-.5cm}
	\label{fig:skewness}
\end{figure}

\textbf{Descriptive analysis.} Descriptive analysis involves examining the data that helps describe, show, or summarize data points. It helps determine patterns or outliers that might emerge, which could lead to further statistical analyses. After cleaning and preprocessing the bug reports from each subject system, we calculate the \emph{cosine similarity} score between each pair of duplicate bug reports using their TF-IDF measures from the textually similar and dissimilar subsets. Then we perform descriptive analysis on these similarity scores and capture five different statistics: Skewness, Kurtosis, Mean, Median, and Standard Deviation. Fig. \ref{fig:skewness} and Table \ref{tab:descriptive-analysis} summarize our descriptive analysis for Eclipse, Firefox, and Mobile systems.

\emph{Skewness} is a measure of symmetry \cite{r35}. Distribution is symmetric if it looks the same on the left and right of the center point \cite{r35}. From Fig. \ref{fig:skewness}, we find the scores of textually similar duplicate bug reports to be positively skewed for Eclipse and Firefox. The positive skewness indicates that a significant number of duplicate pairs are highly similar \cite{r35}. On the other hand, for the textually dissimilar dataset, we found negative skewness for all three datasets, indicating that the cosine similarity measures are very low for most of the textually dissimilar duplicate pairs \cite{r35}. 

\emph{Kurtosis} is a measure of whether the distribution is heavy-tailed or light-tailed relative to a normal distribution \cite{r35}. Distributions with positive kurtosis tend to have heavy tails or outliers, whereas distributions with negative kurtosis tend to have light tails \cite{decarlo1997meaning}. A uniform distribution would be the extreme case. From Fig. \ref{fig:skewness} and Table \ref{tab:descriptive-analysis}, we note that textually dissimilar duplicate pairs have a negative kurtosis for Firefox and Mobile systems. The kurtosis for the Eclipse system is also close to zero. That is, the similarity scores from textually dissimilar duplicate pairs have a lighter tail than the normal distribution. In other words, their similarity scores are mostly centered around the mean value, which is also low.

On the other hand, a similar conclusion can be made for the textually similar duplicate pairs with the Eclipse and Mobile systems. However, it should be noted that the Firefox system contains several times more bug reports than the Eclipse and Mobile systems. The remaining two of three statistics - mean and median - are also several times lower for textually dissimilar duplicate bug reports than their counterparts.
\begin{figure}[htbp]
	\includegraphics[width= 3.5in]{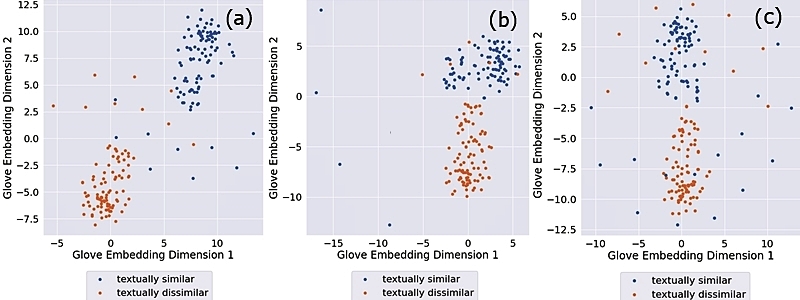}
	\vspace{-.4cm}
	\caption{t-SNE visualization of GloVe embeddings for 100 random samples from both textually similar and dissimilar duplicate bug reports from (a) Eclipse (b) Firefox (c) Mobile system}
	\vspace{-.3cm}
	\label{fig:t-sne}
\end{figure}

\textbf{Embedding analysis.} Word embedding is a frequently used mechanism for detecting duplicate bug reports, representing words as semantically relevant dense real-valued vectors \cite{r62}. While our descriptive analysis above focuses on text-level similarity, we now perform embedding analysis to visualize the semantic differences between textually similar and textually dissimilar duplicate bug reports. We employ t-SNE to visualize high-dimensional data representing embedding vectors in lower dimensions \cite{r32}. It illustrates high-dimensional data (e.g., bug report embeddings) by projecting them into a two-dimensional space. The intra-cluster detail can be observed by measuring the pairwise distances in the higher and lower dimensions spaces \cite{r32}. We collect 100 random bug report pairs from both textually similar and dissimilar datasets. We generated each pair's GloVe embeddings and performed t-SNE to visualize the embedding of each pair. We use the optimal perplexity and iterations (e.g., perplexity = 40, iterations = 7000) and default similarity metric (i.e., cosine similarity) for our visualization.

From Fig. \ref{fig:t-sne}, we see the difference in embedding visualization of textually similar and dissimilar bug reports for all three datasets. In embedding space, textually dissimilar duplicate pairs (orange dots) are clustered in a different dimensional area than the textually similar pairs (blue dots). In all three systems, the embedding positions of textually dissimilar bug reports are in lower coordinates than that of textually similar duplicate bug reports. In particular, the changes are noticeable across the vertical dimension between textually similar and dissimilar duplicate bug reports for Firefox and Mobile. Interestingly, for Eclipse, we see that textually dissimilar duplicate bug reports are at different locations, even across the horizontal dimension. All these differences above suggest that the word semantics of duplicate pairs within the same dataset could be similar but very different from that of other datasets. In other words, the semantics of textually dissimilar duplicate bug reports are noticeably different from that of textually similar bug reports.

\textbf{Manual Analysis.} We chose 100 randomly selected duplicate bug report pairs (50 textually similar + 50 textually dissimilar) from each subject system. Then, we manually analyze 150 pairs of textually similar and 150 pairs of textually dissimilar duplicate bug reports. Ideally, each bug report should have three components -- Expected Behaviour (EB), Observed Behaviour (OB), and Steps to Reproduce (S2R) \cite{bettenburg2008makes}. These components have been used by previous research to reformulate queries during duplicate bug report detection using Information Retrieval \cite{chaparro2019reformulating}. Similarly, we make use of these components from each duplicate pair to understand how textually similar duplicate bug reports and textually dissimilar duplicate bug reports might differ from each other.

First, we go through the \emph{title} and \emph{description} of each bug report and detect the presence of EB, OB, and S2R components in each bug report. Then we analyze the prevalence ratios of these components in both textually similar and textually dissimilar duplicate bug reports. We also calculate the textual similarity between the two bug reports from each pair for each of the three components separately. As a part of manual analysis, we also look for shared terms, keywords, technologies, and overall literary analogies between a duplicate bug report and a master bug report. 
We spent a total of $\approx$25 hours on our manual analysis.

Table \ref{tab:manual} shows the prevalence ratios of all three components from both textually similar and textually dissimilar duplicate bug report pairs. We see that textually dissimilar duplicate bug reports have a higher percentage of missing components. For example, as shown in Table \ref{tab:manual}, 44.61\% and 13\% of textually dissimilar pairs do not contain any steps to reproduce (S2R) and expected behaviors (EB) in their bug reports, whereas such statistics are 13\% and 1\% respectively for the textually similar duplicate bug reports. Such higher ratios of missing components might explain the lower textual similarity between each duplicate pair of their bug reports.

Table \ref{tab:manual} also shows the component-level similarity between two bug reports from each duplicate pair. We see a lower component-level similarity for textually dissimilar duplicate bug reports. For example, on average, two bug reports from each of their pairs are only 38\% and 57\% similar when OB and EB components are considered, whereas such statistics are 83\% and 78\%, respectively, for the textually similar duplicate pairs. Thus, even at the component level, textually dissimilar duplicate bug reports displayed lower similarity ratios. 

In other words, missing components and component-level differences might have led to their overall textual dissimilarity. We also record several qualitative insights during our manual analysis of textually similar and textually dissimilar duplicate bug reports. They are outlined as follows.

(a) \textbf{Shared phrases.} Textually similar duplicate bug reports have more shared phrases (e.g., Bigram, Trigram) than  unique words (e.g., unigram). 
For example, rather than the word "scroll", phrases such as ``horizontal scroll", and ``horizontal scroll installation" are more prevalent in these reports.

(b) \textbf{Components' prevalence.} Observed behaviors (OB) and expected behaviors (EB) are more prevalent in textually similar duplicate bug reports, which could lead to their increased textual similarity. We found up to 90\% overall similarity between two duplicate reports from this category.

(c) \textbf{Missing components.} We observed a higher percentage of missing components in textually dissimilar duplicate bug reports than in textually similar ones. Bug reports with missing components are likely to have lower similarity scores. Although we notice minimal keyword overlaps, the root cause was mostly similar for both bugs from the same duplicate pair.

(d) \textbf{Unique phrases.} Textually dissimilar duplicate bug reports often use completely unique phrases, which could lead to their dissimilarity. We found that although the EB was somewhat similar, the OB and S2R components were written differently for the two bug reports of the same duplicate pair.

\setlength{\arrayrulewidth}{0.1mm}
\setlength{\tabcolsep}{11pt}
\renewcommand{\arraystretch}{1.2}
\begin{table}[!t]
\centering
\caption{Results of Manual Analysis}
\label{tab:manual}
\resizebox{3.5in}{!}{
\begin{threeparttable}
\begin{tabular}{|l|c|c|c|c|c|}
\hline
\multicolumn{1}{|c|}{\textbf{Type}}                                                                        & \multicolumn{1}{c|}{\textbf{EB}} & \multicolumn{1}{c|}{\textbf{OB}} & \multicolumn{1}{c|}{\textbf{S2R}} & \textbf{Overall} \\ \hline
\multicolumn{5}{c}{\textbf{Prevalence Ratio}}\\ 
\hline
\multicolumn{1}{|l|}{\begin{tabular}[c]{@{}l@{}}Textually Similar\\ Duplicate Bug Reports\end{tabular}}    & \multicolumn{1}{c|}{99.00\%}     & \multicolumn{1}{c|}{100.00\%}    & \multicolumn{1}{c|}{87.00\%}      & N/A              \\ \hline
\multicolumn{1}{|l|}{\begin{tabular}[c]{@{}l@{}}Textually Dissimilar\\ Duplicate Bug Reports\end{tabular}} & \multicolumn{1}{c|}{87.05\%}     & \multicolumn{1}{c|}{100.00\%}    & \multicolumn{1}{c|}{55.39\%}      & N/A              \\ \hline
\multicolumn{5}{c}{\textbf{Similarity Ratio}} \\ 
\hline
\multicolumn{1}{|c|}{\begin{tabular}[c]{@{}l@{}}Textually Similar\\ Duplicate Bug Reports\end{tabular}}    & \multicolumn{1}{c|}{78.00\%}     & \multicolumn{1}{c|}{83.00\%}     & \multicolumn{1}{c|}{55.00\%}      & 90.00\%          \\ \hline
\multicolumn{1}{|c|}{\begin{tabular}[c]{@{}l@{}}Textually Dissimilar\\ Duplicate Bug Reports\end{tabular}} & \multicolumn{1}{c|}{56.84\%}     & \multicolumn{1}{c|}{38.20\%}     & \multicolumn{1}{c|}{31.65\%}      & 21.58\%          \\ \hline
\end{tabular}
\end{threeparttable}}
~\\
\textbf{EB}=Expected behaviour, \textbf{OB}=Observed behaviour, \textbf{S2R}=Steps to reproduce
\end{table}

\setlength{\arrayrulewidth}{0.3mm}
\setlength{\tabcolsep}{3pt}
\renewcommand{\arraystretch}{1.6}
\begin{table*}[!t]
    \centering
    \caption{Impact of Domain-Specific Embeddings on Duplicate Bug Report Detection}
    \label{tab: domain-specific embedding}
    \resizebox{6in}{!}{
    \begin{threeparttable}
        \begin{tabular}{|l|cccc|cccc|cccc|}
            \hline
            \multirow{2}{*}{\textbf{Dataset}} & \multicolumn{4}{c|}{\textbf{Eclipse}} & \multicolumn{4}{c|}{\textbf{Firefox}} & \multicolumn{4}{c|}{\textbf{Mobile}} \\ \cline{2-13} 
             & \multicolumn{1}{c|}{\textbf{AUC}} & \multicolumn{1}{c|}{\textbf{Recall}} & \multicolumn{1}{c|}{\textbf{Precision}} & \textbf{F1} & \multicolumn{1}{c|}{\textbf{AUC}} & \multicolumn{1}{c|}{\textbf{Recall}} & \multicolumn{1}{c|}{\textbf{Precision}} & \textbf{F1} & \multicolumn{1}{c|}{\textbf{AUC}} & \multicolumn{1}{c|}{\textbf{Recall}} & \multicolumn{1}{c|}{\textbf{Precision}} & \textbf{F1} \\ \hline
            \multicolumn{12}{c}{\textbf{Pre-trained word embedding only}} \\
            \hline
            \begin{tabular}[c]{@{}c@{}}Test Dataset\end{tabular} & \multicolumn{1}{c|}{61.41} & \multicolumn{1}{c|}{93.00} & \multicolumn{1}{c|}{92.00} & 92.49 & \multicolumn{1}{c|}{64.70} & \multicolumn{1}{c|}{82.00} & \multicolumn{1}{c|}{78.00} & 79.95 & \multicolumn{1}{c|}{84.80} & \multicolumn{1}{c|}{94.00} & \multicolumn{1}{c|}{93.00} & 93.49 \\ \hline
            \begin{tabular}[l]{@{}l@{}}Textually Similar\end{tabular} & \multicolumn{1}{c|}{56.31} & \multicolumn{1}{c|}{55.00} & \multicolumn{1}{c|}{63.00} & 58.73 & \multicolumn{1}{c|}{66.44} & \multicolumn{1}{c|}{52.00} & \multicolumn{1}{c|}{75.00} & 61.41 & \multicolumn{1}{c|}{63.96} & \multicolumn{1}{c|}{72.00} & \multicolumn{1}{c|}{80.00} & 75.79 \\ \hline
            \begin{tabular}[l]{@{}l@{}}Textually Dissimilar\end{tabular} & \multicolumn{1}{c|}{46.74} & \multicolumn{1}{c|}{45.00} & \multicolumn{1}{c|}{51.00} & 47.81 & \multicolumn{1}{c|}{57.06} & \multicolumn{1}{c|}{48.00} & \multicolumn{1}{c|}{59.00} & 52.93 & \multicolumn{1}{c|}{60.99} & \multicolumn{1}{c|}{58.00} & \multicolumn{1}{c|}{76.00} & 65.79 \\ 
            \hline
            \multicolumn{12}{c}{\textbf{Pre-trained word embedding + Oversampling}} \\
            \hline
            \begin{tabular}[c]{@{}l@{}}Textually Similar\end{tabular} & \multicolumn{1}{c|}{56.51} & \multicolumn{1}{c|}{55.00} & \multicolumn{1}{c|}{63.00} & 58.73 & \multicolumn{1}{c|}{66.51} & \multicolumn{1}{c|}{52.00} & \multicolumn{1}{c|}{75.00} & 61.41 & \multicolumn{1}{c|}{63.96} & \multicolumn{1}{c|}{72.00} & \multicolumn{1}{c|}{80.00} & 75.79 \\ \hline
            \begin{tabular}[c]{@{}l@{}}Textually Dissimilar\end{tabular} & \multicolumn{1}{c|}{46.98} & \multicolumn{1}{c|}{47.00} & \multicolumn{1}{c|}{53.00} & 49.82 & \multicolumn{1}{c|}{57.35} & \multicolumn{1}{c|}{49.00} & \multicolumn{1}{c|}{60.00} & 53.94 & \multicolumn{1}{c|}{60.99} & \multicolumn{1}{c|}{58.00} & \multicolumn{1}{c|}{76.00} & 65.79 \\ \hline
            \multicolumn{12}{c}{\textbf{Domain-specific word embedding + Oversampling}} \\
            \hline
            
            \begin{tabular}[c]{@{}l@{}}Textually Similar\end{tabular} & \multicolumn{1}{c|}{56.31} & \multicolumn{1}{c|}{51.00} & \multicolumn{1}{c|}{52.00} & 51.49 & \multicolumn{1}{c|}{56.03} & \multicolumn{1}{c|}{61.00} & \multicolumn{1}{c|}{63.00} & 61.98 & \multicolumn{1}{c|}{64.98} & \multicolumn{1}{c|}{70.00} & \multicolumn{1}{c|}{73.00} & 71.47 \\ \hline
            \begin{tabular}[c]{@{}l@{}}Textually Dissimilar\end{tabular} & \multicolumn{1}{c|}{53.54} & \multicolumn{1}{c|}{51.00} & \multicolumn{1}{c|}{51.00} & 51.00 & \multicolumn{1}{c|}{55.16} & \multicolumn{1}{c|}{60.00} & \multicolumn{1}{c|}{61.00} & 60.50 & \multicolumn{1}{c|}{62.25} & \multicolumn{1}{c|}{69.00} & \multicolumn{1}{c|}{69.00} & 69.00 \\ 
            \hline
        \end{tabular}
    \end{threeparttable}
    }
\end{table*}

\FrameSep.3em
\begin{frshaded}
	\noindent
	\textbf{Summary of RQ$\mathbf{_2}$:} 
	Textually similar and textually dissimilar duplicate bug reports are different in terms of their 
	descriptive statistics (e.g., skewness), underlying semantics (e.g., t-SNE clusters), and prevalence of structural components. In particular, textually dissimilar duplicate bug reports often \emph{miss important components} such as expected behaviors (EB) or steps to reproduce (S2R), which could lead to their textual dissimilarity within each pair.
\end{frshaded} 
\subsection{RQ$_3$: Does domain-specific embedding help improve the detection of textually dissimilar duplicate bug reports?}

From RQ$_1$ and RQ$_2$, we see that textually similar and dissimilar duplicate bug reports could be different in their lexicon, underlying semantics, and structures. Our analysis above also shows that the performance gap between these two sets is the smallest when the Machine Learning approach is used for duplicate bug report detection (RQ$_1$). Machine Learning approaches might be able to capture more contextual information than the other approaches, which could be useful for the detection task \cite{obulesu2018machine, almeida2002predictive, bitvai2015non}. 

In RQ$_1$, we used pre-trained word embeddings from GloVe to train our Siamese CNN model \cite{r14} for duplicate bug report detection. Pre-trained word embedding has proven to be invaluable for improving the performance of various natural language understanding tasks (e.g., text classification \cite{alwehaibi2018comparison}, sentiment analysis \cite{rezaeinia2019sentiment}). However, GloVe has been pre-trained on natural language texts (e.g., Wikipedia) \cite{r54}, which might not be relevant to the texts from bug reports. Thus, we use domain-specific embedding to retrain our Siamese CNN model. We set the max token size to 20,000 and the embedding dimension to 100, which are similar to the parameters used in RQ$_1$. We generate the embedding matrix with Skip-gram algorithm \cite{nooralahzadeh2018evaluation}, use the whole dataset of 92,854 bug reports, and apply the same deep learning architecture to Siamese CNN, as used in RQ$_1$ \cite{r14}.

Datasets constructed from bug-tracking systems are often heavily imbalanced. The number of duplicate bug reports is considerably smaller than that of non-duplicate bug reports \cite{aggarwal2017detecting}. Hence, we use \emph{oversampling} \cite{yap2014application}
to handle the data imbalance problem during model training \cite{yap2014application}. 
To the best of our knowledge, the original work \cite{r14} did not use sampling in their Siamese CNN model. However, for an in-depth investigation, we replicate another variant of the original DL-based model \cite{r14} applying oversampling.   
Table \ref{tab: domain-specific embedding} summarizes the experimental results of our DL-based model for three different scenarios: \emph{pre-trained embedding only} (original work \cite{r14}), \emph{pre-trained embedding + oversampling}, and \emph{domain-specific embedding + oversampling}. When compared between these model scenarios-- pre-trained embedding + oversampling, and domain-specific embedding + oversampling, we see the noticeable impact of domain-specific embeddings on duplicate bug report detection. 

From Table \ref{tab: domain-specific embedding}, we see that, in terms of F1-measure, the model's performance with textually dissimilar duplicate bug reports has increased by 1.18\%  for Eclipse, 6.56\% for Firefox, and 3.21\% for Mobile. Furthermore, the AUC improved by 6.56\% for Eclipse and remained comparable for the other two systems. 
Thus, domain-specific embeddings have a positive impact on detecting textually dissimilar duplicate bug reports. However, they have mostly negative impacts, except in a few cases, on detecting textually similar bug reports. From Table \ref{tab: domain-specific embedding}, we see that the model's F1-measure decreased by 7.24\% for Eclipse and 4.32\% for the Mobile system. Furthermore, the AUC decreased by 10.48\% for the Mozilla system. Thus, while domain-specific embeddings have the potential to tackle the challenge of textual dissimilarity, they have a mixed impact on detecting duplicate bug reports.    

\FrameSep.3em
\begin{frshaded}
	\noindent
	\textbf{Summary of RQ$\mathbf{_3}$:} The use of domain-specific embeddings (e.g., trained on bug reports) improves our model's performance for textually dissimilar duplicate bug reports (e.g., up to \textbf{6.56}\% in F1-measure for Firefox). However, these embeddings have either negligible or negative impacts on detecting textually similar duplicate bug reports.
\end{frshaded} 
\section{Threats to Validity} \label{sec:threats}
We identify a few threats to the validity of our findings. In this section, we discuss these threats and the necessary steps taken to mitigate them as follows.

\textbf{Threats to internal validity} relate to experimental errors and human biases \cite{tian2014automated}. Traditional bug tracking systems (e.g., Bugzilla) have thousands of reports whose quality cannot be guaranteed, which could be a source of threat. Bug reports often contain poor, insufficient, missing, or even inaccurate information \cite{gupta2021systematic}. To address the issue, we apply standard natural language preprocessing and token threshold to them and also check for missing features in each bug report. Another potential source of threat could be the replication and reproduction of existing work. The replication package was unavailable for the BM25 and Siamese CNN models, and we had to re-implement them. However, we did it carefully using standard libraries and corresponding papers, tuned the parameters, and reported their best results.

We use TF-IDF and cosine similarity to determine the textual similarity between any two duplicate bug reports. TF-IDF and cosine similarity have been frequently used to determine the textual similarity between two documents for the last 50 years \cite{jones1972statistical}. Besides, we also used N-gram-based similarity and quartile analysis to systematically separate the textually similar and textually dissimilar duplicate bug reports (Section \ref{sec:methodology}), and report the detailed similarity measures for replication (see Table \ref{tab:similarity_score} Thus, the threats concerning similarity calculation and construction of two bug report groups might be mitigated. 

\textbf{Threats to conclusion validity.} The observations from our study and the conclusions we drew from them could be a source of threat to conclusion validity \cite{garcia2012statistical}. In this research, we answer three research questions using 92,854 bug reports from three different subject systems and re-implementing three existing techniques. We use appropriate statistical tests (e.g. Wilcoxon Signed Rank) and report the test details (e.g., p-value, Cliff's delta) to draw any conclusion. Thus, such threats might also be mitigated.

\textbf{Threats to construct validity} relate to the use of appropriate performance metrics. We evaluate BM25 and LDA+GloVE techniques with Recall-rate@K and Machine Learning model with AUC, precision, recall, and F1-measure, which have been used in their corresponding papers and the relevant literature \cite{gupta2021systematic}. 
Thus, such threats might also be mitigated.
\section{Related Work}\label{sec:related-work}
\textbf{Information Retrieval (IR).}
IR approaches rely on the textual overlap between query bug reports and candidate bug reports for duplicate detection. \citet{r11} first use a simple approach
namely Bag of Words (BOW), to tally the frequency of words and then use BOW-model to detect duplicate bug reports. They determine the similarity between two bug reports
using cosine, Jaccard, and dice similarity measures. However, the BOW-based approach could be biased towards large documents and might not be able to capture the semantics of a bug report precisely \cite{wu2010semantics}. \citet{r21} later improved this technique using TF-IDF \cite{aizawa2003information} and quantify the similarity of two document vectors. 

Later, they used BM25 \cite{robertson2009probabilistic} as a traditional IR-based model for duplicate bug report detection. \citet{aggarwal2017detecting} demonstrate that BM25F, an improvement of BM25, is more suitable for weighting words in diverse domains. Like us, they also use domain-specific, categorical, and textual features.\cite{sureka2010detecting} use n-gram models for textual similarity calculation during duplicate bug report detection. In particular, they focus on character-level language models rather than word-level ones. We also use n-gram-based similarity to separate textually similar and textually dissimilar duplicate bug reports.
 
\citet{chaparro2019reformulating} use three strategies to reformulate a query bug report and use Information Retrieval to detect duplicate bug reports. Later, \citet{cooper2021takes} propose a duplicate bug report detection for video-based bug reports where they make use of text retrieval and computer vision methods. Since we focus on textual bug reports, their work might not be a great fit. In our research, we thus use BM25, a popular IR baseline, to investigate the impacts of textual dissimilarity on duplicate bug report detection. 

\textbf{Topic Modeling.}
IR-based approaches might suffer from Vocabulary Mismatch Problems \cite{vocaprob}. Topic Modeling has the potential to tackle such problems concerning textual similarity calculation \cite{wallach2006topic}. \citet{alipour2013contextual} employ the Latent Dirichlet allocation (LDA) model to capture contextual information from history (e.g., prior knowledge on software quality) and leverage the information in duplicate bug report detection. \citet{aggarwal2017detecting} capture domain-specific contextual information to improve duplicate bug report detection. 

\citet{nguyen2012duplicate} combine IR and topic-based features to improve duplicate bug report detection. Recently \citet{r13} propose a hybrid model that combines the Topic Modeling (e.g., LDA) with pre-trained word embedding (e.g., GloVe) for duplicate bug report detection.
We replicate their technique carefully for our experiments, and detailed results can be found in Table \ref{tab:ir-whole-dataset}.
In another research, \citet{budhiraja2018lwe} combines Latent Dirichlet Allocation (LDA) and domain-specific word embeddings. Similarly, we leverage domain-specific embedding to counteract the impact of textual dissimilarity in duplicate bug report detection (RQ$_3$).

\textbf{Machine Learning and Deep Learning.} Unlike the above two methodologies, Machine Learning can detect non-linear relationships between any two bug reports for duplicate detection \cite{obulesu2018machine, almeida2002predictive, bitvai2015non}. \citet{r29} first used a Support Vector Machine (SVM) to design a discriminative model for detecting duplicate bug reports. However, their approach lacks rigorous validation. \citet{r45} design several models using K-NN, Linear SVM, RBF, Decision Tree, Random Forest, and Naive Bayes to classify duplicate bug reports. 

\citet{r14} used the Siamese variations of CNN and RNN to design a deep-learning model for duplicate bug report detection. We replicate their work for our experiments.
\citet{r46} incorporate the attention mechanism into the Siamese network for semantic and context-based embedding. \citet{r47} propose an architecture, namely DBR-CNN, where CNN is used to encode the textual data and logistic regression to classify each pair of bug reports as either \emph{duplicate} or \emph{non-duplicate}. 
\citet{haering2021automatically} employ DistilBERT, a context-sensitive embedding technique using BERT and deep matching for duplicate bug report detection.

To summarize, we replicate three existing techniques on duplicate bug report detection from Information Retrieval, Topic-Modeling, and Deep Learning. Then we conduct experiments using a total of 92K bug reports to better understand the impacts of textual dissimilarity on duplicate bug report detection.
To our best knowledge, this is the first attempt to comprehensively understand the impacts of textual dissimilarity on duplicate bug detection, which makes our work \emph{novel}.

\section{Conclusion \& Future Work} \label{sec:conclusion}
Automated detection of duplicate bug reports has been an active research topic for over a decade. However, existing approaches might not be sufficient to detect textually dissimilar but duplicate bug reports. In this paper, we thus perform a large-scale empirical study using 92K bug reports from three open-source systems to better understand the challenges of textual dissimilarity in duplicate bug report detection. First, we empirically demonstrate that existing techniques perform poorly in detecting textually dissimilar duplicate bug reports. Second, we found that textually dissimilar duplicates often miss important components (e.g., steps to reproduce), which could lead to their textual dissimilarity within the same pair. Finally, inspired by the earlier findings, we apply domain-specific embedding to duplicate bug report detection, which provided mixed results. All these findings above warrant further investigation and more effective solutions for detecting textually dissimilar duplicate bug reports. Future work can focus on complementing these reports with relevant information from other sources (e.g., version control history).

\bibliographystyle{plainnat}
\bibliography{ref1.bib}
\end{document}